\RequirePackage{fix-cm}
\documentclass[twocolumn]{svjour3}          
\smartqed  
\usepackage{graphicx}
\newcommand{\cf}{\textit{cf.}}
\newcommand{\eg}{\textit{e.g.}}
\newcommand{\etal}{\textit{et al.}}
\newcommand{\etc}{\textit{etc.}}
\newcommand{\ie}{\textit{i.e.}}

\newcommand{\ttl}[1]{\textit{{#1}}}

\journalname{KI -- K\"unstliche Intelligenz}
\begin{document}

\title{Ontologies for the Virtual Materials Marketplace\thanks{The Virtual Materials Marketplace (VIMMP) project is funded from the European Union's Horizon 2020 research and innovation programme under grant agreement no.\ 760907.}}

\titlerunning{Ontologies for VIMMP}  

\author{Martin Thomas Horsch \and Silvia Chiacchiera \and Michael A.\ Seaton \and Ilian T.\ Todorov \and Karel \v{S}indelka \and Martin L\'isal \and Barbara Andreon \and Esteban Bayro Kaiser \and Gabriele Mogni \and Gerhard Goldbeck \and Ralf Kunze \and Georg Summer \and Andreas Fiseni \and Hauke Br\"uning \and Peter Schiffels \and Welchy Leite Cavalcanti}

\authorrunning{M.\ T.\ Horsch \etal} 

\institute{
M.\ T.\ Horsch, S.\ Chiacchiera, M.\ A.\ Seaton, I.\ T.\ Todorov \at
   STFC Daresbury Laboratory, Keckwick Ln, Daresbury, Cheshire WA4 4AD, UK \\
   Phone: +44 1925 86 4318,
   fax: +44 1925 60 3100 \\
   \email{\{martin.horsch, silvia.chiacchiera, michael.seaton, ilian.todorov\}@stfc.ac.uk}
\and
K.\ \v{S}indelka, M.\ L\'isal \at
   Department of Molecular and Mesoscopic Modelling, Institute of Chemical Process Fundamentals of the Czech Academy of Sciences, v.v.i., Rozvojov\'a 135/1, 165 02 Prague 6-Suchdol, Czech Republic \\
   Phone: +420 220 390 301,
   fax: +420 220 920 661 \\
   \email{\{sindelka, lisal\}@icpf.cas.cz}
\and
B.\ Andreon, E.\ Bayro Kaiser \at
   WearHealth UG, Fahrenheitstr.\ 1, 28359 Bremen, Germany \\
   Phone: +49 176 6175 9285,
   fax: +49 421 218 64047 \\
   \email{\{barbara, esteban\}@wearhealth.com}
\and
G.\ Mogni, G.\ Goldbeck \at
   Goldbeck Consulting Ltd, St John's Innovation Centre, Cowley Rd, Cambridge CB4~0WS, UK \\
   Phone: +44 1223 853201 \\
   \email{\{gabriele, gerhard\}@goldbeck-consulting.com}
\and
R.\ Kunze, G.\ Summer, A.\ Fiseni \at
   Osthus GmbH, Eisenbahnweg 9--11, 52068 Aachen, Germany \\
   Phone: +49 241 943140,
   fax: +49 241 9431419 \\
   \email{\{ralf.kunze, georg.summer, andreas.fiseni\}@ost\-hus.com}
\and
H.\ Br\"uning, P.\ Schiffels, W.\ L.\ Cavalcanti \at
   Fraunhofer Institute for Manufacturing Technology and Advanced Materials, Wiener Str.~12, 28359 Bremen, Germany \\
   Phone: +49 421 22460,
   fax: +49 421 2246 300 \\
   \email{\{hauke.bruening, peter.schiffels, welchy.leite.caval\-canti\}@ifam.fraunhofer.de}
}

\date{Received: date / Accepted: date}

\maketitle

\begin{abstract}
The Virtual Materials Marketplace (VIMMP) project, which develops
an open platform for providing and accessing services related to materials
modelling, is presented with a focus on its ontology development and data
technology aspects. Within VIMMP, a system of marketplace-level ontologies
is developed to characterize services, models, and interactions between users;
the European Materials and Modelling Ontology is employed as a top-level ontology.
The ontologies are used to annotate data that are stored in the ZONTAL Space component of VIMMP
and to support the ingest and retrieval of data and metadata at the VIMMP marketplace frontend.
\end{abstract}

\section{Introduction}
\label{intro}

In mechanical and process engineering, the digitalization of industrial manufacturing and process design is strongly tied to innovations in process data technology~\cite{KHKR14,MB14,FJBBVH19,JADVBBMKH20}. In this context, semantic technology (\eg, data management based on ontologies) is crucial as it facilitates the integration of data with a diverse and heterogeneous provenance, including experiment, simulation, and machine-learning based surrogate models, into coherent frameworks~\cite{JADVBBMKH20,AB4WBHKH14,FGKJKQSSMACBJ19}. By combining multiple source data sets, repositories, or research data infrastructures, simulation results can be evaluated and assessed for consistency~\cite{STVH19}. As a consequence of these developments, the market for services related to modelling and simulation of thermodynamic and mechanical properties is expected to increase significantly. For the period between 2017 and 2021, the annual growth rate for computer aided engineering, which includes computational molecular engineering in the field of fluid thermodynamics~\cite{HNBCSCELNSSTVC20} and integrated computational materials engineering in the field of solid mechanics~\cite{Schmitz16}, is predicted to exceed 10\%~\cite{GS18}; supporting the long-term sustainability of these developments, the Virtual Materials Marketplace (VIMMP), an open two-sided marketplace platform for services in materials modelling, is presently under development within an eponymous Horizon 2020 innovation action~\cite{VIMMP19}. The portfolio of models and simulation methods includes electronic, atomistic, mesoscopic, and continuum approaches, and all thermodynamic and mechanical properties of materials are considered, including properties of homogeneous fluid systems and multiphase systems.

Internally, VIMMP uses ontologies as a part of its approach to data management, underlying, in particular, the interactions with users at its frontend. Additionally, ontologies are used to facilitate semantic interoperability with external platforms providing services that may be offered via VIMMP as well as related data and metadata. For this purpose, VIMMP and a series of related projects contribute to the activities of the European Materials Modelling Council (EMMC), an association with a focus on reliability, interoperability, standardization, and industrial usability of methods in computational molecular engineering~\cite{EMMC18,EMMC19b,EMMC19c}. The present work summarizes the ongoing efforts and objectives of the VIMMP project concerning ontology development and its relation to data management. It is structured as follows: In Section \ref{sec:marketplace-level}, we briefly describe the marketplace services and correspondigly introduce the related marketplace-level ontologies; then, Section \ref{sec:data} summarizes the role of the ontologies in data management, including data ingest and data retrieval.
In Section~\ref{sec:emmo}, it is discussed how the present work can support the efforts of the EMMC to enable inter-platform interoperability by alignment with the European Materials and Modelling Ontology (EMMO), a community-governed top-level ontology for modelling and characterizing materials~\cite{EMMC19b,HCBSMGG20}, before drawing conclusions in Section~\ref{sec:conclusion}.

\section{Marketplace-level ontologies and interoperability}
\label{sec:marketplace-level}

The services provided on the VIMMP platform will target a heterogeneous
community of users with diverse backgrounds
and needs. Furthermore, they will need to cover a broad range of facets of materials modelling;
this includes training (both in the form of training material and events),
a catalogue of experts, translation (from an industrial problem to a modelling
solution), open simulation platforms, software tools, workflows, model parameters, \etc{}
For a platform design that involves interactions between many components,
developed both internally and externally,
semantic interoperability of these components can significantly
enhance the functionality and performance, \eg, by permitting
simultaneous querying of multiple repositories that have different
internal data representations. Deeper forms of
interoperability are conceivable on this basis, such as an interchangeability of
software tools and portability of models and simulation workflows
between multiple simulation platforms.
Ontologies define the ``common language'' semantics used for this purpose.

Accordingly, an ecosystem of ontologies is developed within VIMMP; these
ontologies are expressed in OWL2 using TTL notation.
This system will not be frozen: There will be a policy to allow users and providers to request
the addition of categories or to integrate subdomain-specific ontologies.
Any provider will have the possibility to choose the depth at which any
provided services and tools implement the proposed common
semantics: The deeper the adherence, the deeper the interoperability with other services.
The system of ontologies from VIMMP and its connection to externally developed and pre-existing semantic assets is shown in Fig.~\ref{fig:architecture}. Therein, if an ontology references concepts (\eg, class definitions) from another semantic asset, an arrow is shown, leading from the component of the architecture where the concept is used to the component where it is defined; accordingly, \eg, class definitions from the European Virtual Marketplace Ontology~(EVMPO)~\cite{HCBSMGG20} are employed, and hence pointed to, by all other ontologies developed in VIMMP.
At the marketplace level, focusing on service, platform, and simulation interoperability,
this includes eight components which are described below:

%
%

\begin{figure}
\centering
\includegraphics[width=0.4833\textwidth]{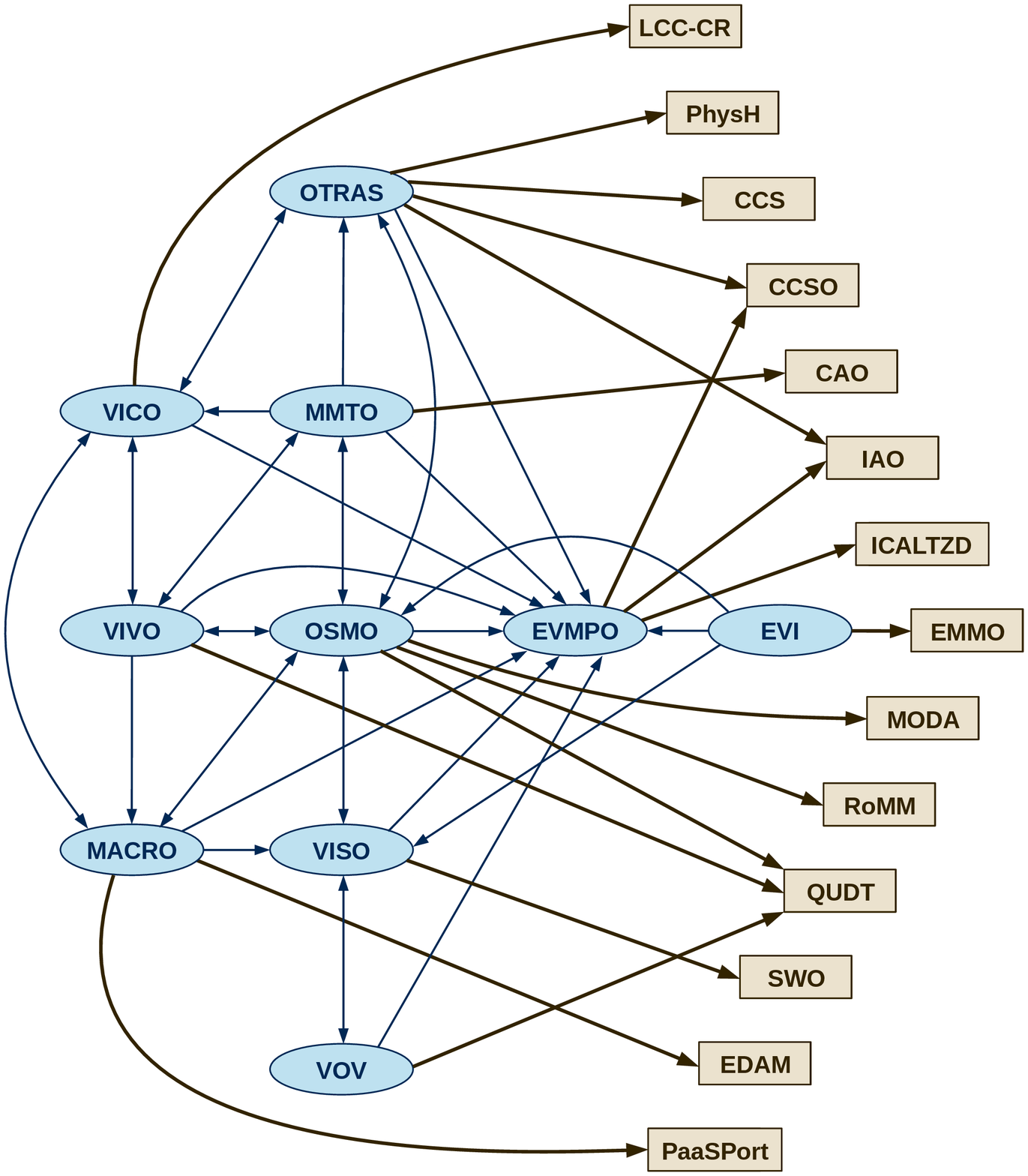}
\caption{Ontologies from VIMMP (ellipses) and referenced external semantic assets (rectangles).}
\label{fig:architecture}
\end{figure}

The Marketplace-Accessible Computational Re\-sour\-ce Ontology (MACRO) deals with data and hardware related resources and infrastructures. In particular, MA\-CRO contains classes and individuals representing the file formats that are expected to occur on the VIMMP marketplace platform, which it connects to the EDAM ontology~\cite{IKJBUMMLPR13}. The PaaSPort ontology~\cite{BSGKMV18} is used to describe \textit{platforms as a service}~(PaaS).

Translators, \ie, facilitators of industrial use cases in materials modelling and simulation, and the associated translation processes, as described by the EMMC Translators' Guide~\cite{EMMC18}, are within the domain of the Materials Modelling Translation Ontology~(MMTO); this includes an ontology-level formalization of the EM\-MC Translation Case Template~\cite{EMMC19c}. The Currency A\-mount Ontology~(CAO), which is a component of the Financial Industry Business Ontology~\cite{Bennett14}, is employed to specify costs and benefits for the associated business cases.

The Ontology for Simulation, Modelling, and Optimization~(OSMO) is the ontology version of the pre-existing semiformal simulation workflow graph standard MODA (``model data'')~\cite{CEN18}; in OSMO, simulation workflows are described on the basis of MODA and the Review of Materials Modelling (RoMM), cf.~de Baas~\cite{DeBaas17}, and the logical data transfer between elements of a simulation workflow is characterized~\cite{HNBCSCELNSSTVC20}. In particular, this includes the classification of widespread types of physical equations at the electronic, atomistic, mesoscopic, and continuum levels from RoMM~\cite{DeBaas17}.

The Ontology for Training Services~(OTRAS) includes a taxonomy of topics in materials modelling, also permitting the inclusion of topics from CCS~\cite{ACM12} and PhysH~\cite{APS19}, and a formalism by which learning outcomes and expert competencies can be described. For information on training courses, syllabi, \etc, the Course Curriculum and Syllabus Ontology~(CCSO) is employed~\cite{KKAK18}. 

The VIMMP Communication Ontology~(VICO) covers metadata on messages exchanged at the virtual marketplace platform and participants that interact at the platform, including end users, model providers, software owners, translators, \etc{} Through the LCC-CR ontology, VICO incorporates the ISO 3166 standard for referring to countries and regions~\cite{Celko10}.

On the basis of the VIMMP Software Ontology~(VI\-SO), features and properties, including
licensing aspects, can be specified for software packages that
are offered on the marketplace. VISO focuses in particular on the software
capabilities, both at the modelling and numerical level, and includes three
subdomain-specific modules for electronic (\ttl{viso-el}), atomistic/mesoscopic (\ttl{viso-am}), and
continuum (\ttl{viso-co}) models.
Some concepts from VISO, as software interfaces, include references to the pre-existing Software
Ontology (SWO), \cf~Malone \etal~\cite{MBLIHPS14}.

The VIMMP Validation Ontology (VIVO) addresses model, solver, and processor errors, assessments of computational resource requirements and benchmarking as well as customer feedback to be provided for transactions at the virtual marketplace platform.

The relation of physical properties to models, solvers, and the variables occurring in them is in the scope of the VIMMP Ontology of Variables (VOV); it is closely related to VISO and OSMO. Concerning physical properties and units, OSMO, VIVO, and VOV refer to the QUDT system of ontologies~\cite{ZLZP17}.

\section{Data ingest, management, and retrieval}
\label{sec:data}

For internal data management, the Allotrope Data Format (ADF) is employed by VIMMP as a data standard. This format is developed by the Allotrope Consortium, coordinated by the VIMMP project partner Osthus GmbH as a framework architect~\cite{CU15}. ADF is built on the well-established HDF5 file specification for storage of data in a binary format, within which acquired data are stored in one or more \textit{data cubes}. Metadata are stored as \textit{data description} triples using a \textit{RDF data model}, aligned to the VIMMP ontologies; further metadata are provided to describe the \textit{data cube} and \textit{data package} layers, all based on semantic web and linked data concepts and the appropriate W3C specifications. In particular, VIMMP takes advantage of ADF by using ZONTAL Space, which is also developed by Osthus GmbH, as a central storage component. ZONTAL Space is a data lifecycle management system based on the ISO 14721:2012 standard for Open Archival Information Systems~\cite{ISO12}. In this way, syntactic and semantic interoperability are achieved coherently, laying an extensible foundation for connecting VIMMP to state-of-the-art data analytics, research data infrastructures, and big data platforms.

Accordingly, the ontologies are used for purposes of data management at the backend.
Moreover, they are employed to guide the data
ingest, \eg, to create profiles for agents (using VICO) and to document
computational services (using MACRO), software (using VISO), and models (using OSMO)
as well as training materials and events (using OTRAS), and
to assist at data retrieval by search, browsing, and similar functionalities
at the frontend, \ie, the user interface.
Since, initially, services will be predominantly provided by VIMMP consortial partners,
it will be a priority during the launch phase to populate the platform with external service providers.
To ensure that the business models of service providers and the VIMMP marketplace are viable,
end users need to be attracted as clients, and they
need to be enabled to find the services and providers that they need without unnecessary complications.
Users will indirectly be in contact with the ontologies via the available keywords as well as search criteria and outcomes.

\begin{figure*}
\centering
\includegraphics[width=0.8667\textwidth]{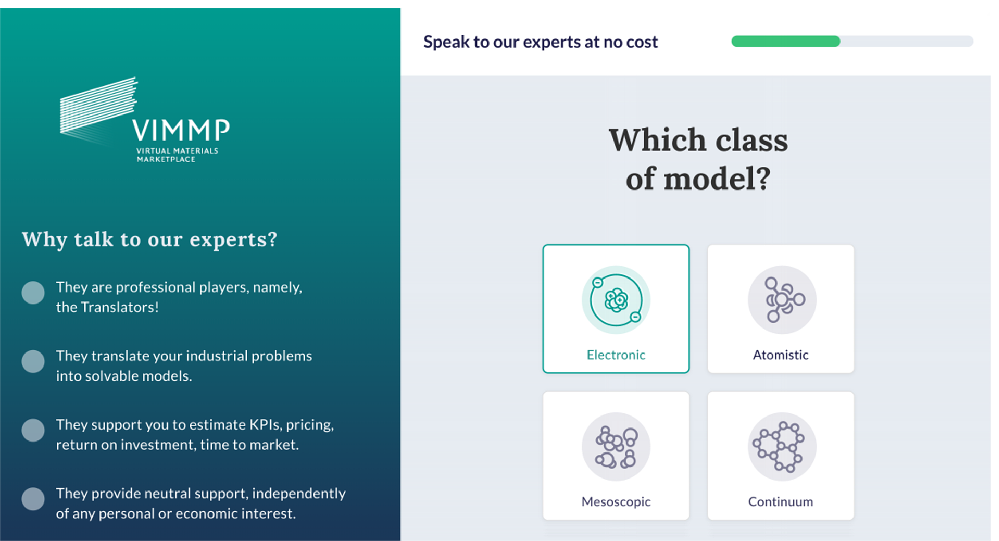}
\caption{End-user data ingest for translation service data retrieval at the \textit{translation router} app of the VIMMP marketplace frontend prototype.}
\label{fig:data-ingest}
\end{figure*}

Beside accessibility of the ontologies, it is therefore essential to complement them
by an infrastructure that makes data ingest and data retrieval as user friendly as possible.
In Fig.~\ref{fig:data-ingest}, a snapshot from the prototype \textit{translation router}
is shown; this frontend app allows a prospective end user to provide relatively generic
initial information \eg, on the type of problem and the preferred modelling approach.
Eventually, this is matched to one or multiple
profiles of suitable translators, \ie, professional engineers or academics
who facilitate industrial applications of materials modelling;
this is the established meaning of the
word \textit{translator} in the EMMC community~\cite{EMMC18,EMMC19c}.
The retrieval of these profiles occurs by querying for topic codes, defined in the ontology OTRAS,
which need to agree with topics in which the respective providers of translation
services have indicated their interest.

\section{Top level and fundamental categories}
\label{sec:emmo}

The European Materials and Modelling Ontology (EM\-MO), developed within the EMMC-CSA project~\cite{EMMC19b}, is employed as a top-level ontology within VIMMP. The EMMO is work in progress (hence, alignment of the VIMMP ontologies with the EMMO is also work in progress); to support the community uptake, a pre-release version was made available to VIMMP and other projects that contribute to EMMC efforts. The most recent developments can be followed at a github repository~\cite{EMMC19b}.
The EMMO aims at supporting semantic interoperability by involving contributors from all fields of materials modelling in a development effort that is guided
by a philosophical approach and basic design that permits to address any problem that could arise across a variety of applications and use cases in the future -- not limited to modelling and simulation of physical properties of materials, but also concerning materials characterization, experimental data, and the empirical observation of physical systems. The two major aspects of this approach are mereotopology, following Varzi~\cite{Varzi96}, and semiotics, following Peirce~\cite{Peirce91}.

Below the level of the EMMO, the EVMPO defines concepts that are constitutive to the paradigm of a virtual marketplace. The EVMPO was jointly agreed between VIMMP and the MarketPlace project~\cite{Marketplace19} and is intended to serve as a basis for further future work toward platform interoperability. At the greatest level of abstraction, the EVMPO splits this paradigmatic semantic space into eleven branches corresponding with ``fundamental paradigmatic categories'' as topmost clas\-ses: (1) \ttl{assessment}, (2) \ttl{calendar\_event}, (3) \ttl{communication}, (4) \ttl{information\_content\_entity}~\cite{Ceusters12}, (5) \ttl{infrastructure}, (6) \ttl{interpreter}, (7) \ttl{material}, (8) \ttl{model}, (9) \ttl{process}, (10) \ttl{product}, and (11) \ttl{property}. These categories are related to corresponding entities from the EMMO, where available, or to the appropriate superclasses from the EMMO through a module for EMMO-VIMMP Integration (EVI). Additionally, the definition of \ttl{calendar\_event} points to the W3C iCalendar ontology (ICALTZD), \cf~Connolly and Miller~\cite{CM05}, while \ttl{information\_content\_entity} is taken from the Information Artifact Ontology (IAO) \cite{Ceusters12}, cf.~Fig.~\ref{fig:architecture}.

\section{Conclusion}
\label{sec:conclusion}

Ontology development in the VIMMP project serves multiple objectives: At the backend, it is intended to support data management. At the frontend, to facilitate a user-friendly data ingest and data retrieval, the ontologies and their role need to be hidden from platform users: Concepts need to be referred to in such a way that the users, not the ontology developers, understand them; on the other hand, users who wish to engage with these aspects of the platform must have the opportunity to access them and suggest extensions and modifications. Concerning a system of platforms and resources embedded within the semantic web, the purpose of the ontologies consists in facilitating service interoperability. The orientation toward this objective leads to joint developments with other projects, coordinated and supported by discussions within organizations such as the EMMC and the Research Data Alliance. Additionally, the desire for a consistent formalization of the employed concepts at the highest possible level of abstraction, arising within these organizations, motivates the use of a top-level ontology.

The domains of interest corresponding to these goals can occasionally be
remote from each other;
considerations at the highest level of abstraction (\eg, general merotopology and semiosis) arise independently from technical platform requirements, but they need to be addressed together, considering that
the EMMO is the common top-level ontology for a variety of infrastructures that are presently under construction, including platforms that are expected to be connected to VIMMP. Accordingly, the present architecture of semantic assets was designed to ensure that the EMMO, which is employed at the top level, is supplemented by the EVMPO which contains the high-level classes that are most essential for describing the resources that are relevant to VIMMP and comparable platforms. The EVMPO and the EMMO are connected by the EVI module which is specifically designed for this purpose. Ongoing work on ontology alignment~\cite{HCBSMGG20} will extend this module to create additional connections between marketplace-level ontologies and the EMMO.


\end{document}